\documentclass[twocolumn,aps,prl,showpacs]{revtex4}             

\usepackage{graphics,dcolumn}
\usepackage{bm}
\usepackage[fleqn]{amsmath}
\usepackage{hyperref}
\hypersetup{
     colorlinks=true,       
    linkcolor=blue,          
    citecolor=blue,        
    filecolor=magenta,      
    urlcolor=cyan           
}
\usepackage{exscale,relsize}
\usepackage{epsfig}
\usepackage{amssymb}
\usepackage{amsmath}
\usepackage{euscript}
\usepackage{ulem}
\usepackage{float}
\usepackage{tikz}
\usetikzlibrary{snakes}
\usetikzlibrary{decorations.pathmorphing}
\usetikzlibrary{patterns}
\usepackage{bbm}
\usepackage{footmisc}
\usepackage{epsfig}
\usepackage{amssymb}
\usepackage{amsmath}
\usepackage{euscript}
\usepackage{float}
 \usepackage{amsthm}

\newcommand{\be}{\begin{equation}}
\newcommand{\e}{\end{equation}}

\newcommand{\beml}{\begin{subequations}}
\newcommand{\eml}{\end{subequations}}
\newcommand{\beq}{\begin{eqnarray}}
\newcommand{\eq}{\end{eqnarray}}
\newcommand{\ba}{\begin{array}}
\newcommand{\ea}{\end{array}}

\newcommand{\lt}{\left}
\newcommand{\rt}{\right}
\newcommand{\n}{\nonumber}

\newcommand{\ra}{\rangle}

\begin{document}

\date{\today}

\title{Universal entanglement decay in atmospheric turbulence}

\author{Nina D. Leonhard\footnote{Present address: Fraunhofer Institute for Applied Optics and Precision Engineering, Albert-Einstein-Str. 7, 07745 Jena, Germany}}
\affiliation{Physikalisches Institut, Albert-Ludwigs-Universit\"at Freiburg, Hermann-Herder-Str. 3,
D-79104 Freiburg, Germany}

\author{Vyacheslav N. Shatokhin}

\affiliation{Physikalisches Institut, Albert-Ludwigs-Universit\"at Freiburg, Hermann-Herder-Str. 3,
D-79104 Freiburg, Germany}

\author{Andreas Buchleitner}

\address{Physikalisches Institut, Albert-Ludwigs-Universit\"at Freiburg, Hermann-Herder-Str. 3,
D-79104 Freiburg, Germany}

\begin{abstract} 
We consider the propagation of two photonic qubits, initially maximally entangled in their orbital angular momenta (OAM), 
across a turbulent atmosphere. By introducing the {\it phase correlation length} of an OAM beam, we show that the entanglement of OAM photons exhibits a universal exponential decay under turbulence.
\end{abstract}

\pacs{
03.67.Hk, 
42.50.Tx,
42.68.Bz
}

\maketitle

The ability of photons carrying orbital angular momentum (OAM) to encode quantum states in a high dimensional Hilbert space makes them potentially very useful for quantum 
information purposes \cite{leach02,vaziri03,zeilinger06,Molina-Terriza:2007zr,Nagali:2009qf,pors11,leach12,Fickler02112012}, among which  
free space quantum communication is one of the most promising future applications. Proof-of-principle experiments have shown a significant increase of the classical channel capacity using OAM multiplexing 
\cite{Wang:2012uq}. However, before the realization of (quantum) OAM multiplexing in free space, reliable transport of OAM photons through atmospheric turbulence has to be accomplished \cite{Torres:2012zr}. 

The transmission of photons carrying OAM across turbulence is challenging because the intrinsic refractive index fluctuations associated with turbulence distort the photons' wave fronts \cite{paterson05} that encode quantum information, resulting in a deterioration thereof. In recent years, experimental \cite{Pors:11,Malik:12,alpha13,PhysRevA.88.053836,rodenburg14} and theoretical \cite{raymer06,gopaul07,roux11,Sheng:12,todd_brun13} efforts have been dedicated to clarify and to partially mitigate \cite{PhysRevA.88.053836,rodenburg14,todd_brun13}
the impact of turbulence on single and entangled OAM photons. Despite some progress, there are still  fundamental open issues concerning the behavior of OAM photons in turbulence -- one of them being the description of the OAM photons entanglement evolution. 

A crucial difficulty for the quantitative description of the latter stems from the fact that optical inhomogeneities induced by density fluctuations of the air induce coupling of the initially excited,
finite number of OAM modes to all modes of the infinite-dimensional OAM space. 
Therefore, theoretical methods to treat the open system entanglement evolution of finite-dimensional quantum systems \cite{Mintert05b,tiersch08} are not directly applicable. Indeed, to deal with a necessarily finite-dimensional output state upon detection, we need to 
truncate the Hilbert space, what unavoidably leads to a loss of norm. Therefore, it is more appropriate to use the tools for entanglement characterization of decaying states \cite{hiesmayr03}. 

In this letter we report on some new insights on the entanglement evolution in weak turbulence. Specifically we consider the example of the simplest decaying OAM state -- a maximally entangled OAM qubit, that is, a biphoton whose wave 
front represents a superposition of two spatial Laguerre-Gaussian (LG) modes. We introduce the {\it phase correlation length} $\xi(l)$ -- an inherent property of an LG beam with angular momentum $l$, reflecting its complex spatial structure -- and show that, remarkably, 
 the entanglement exhibits a universal exponential decay as a function of $\xi(l)/r_0$, vanishing at $\xi(l)/r_0\approx 1$, where $r_0$ is the turbulence correlation length defining the characteristic scale of the turbulent `granularity'  over a given propagation distance $L$ also called Fried parameter. If $\xi(l)\ll r_0$, the turbulent atmosphere appears as a homogeneous medium to the OAM biphoton, and its spatial entanglement remains high. 
As $r_0$ approaches $\xi(l)$, the phase errors become sufficiently large to destroy the wave front structure, and the entanglement vanishes.

\begin{figure}
\includegraphics[width=8cm]{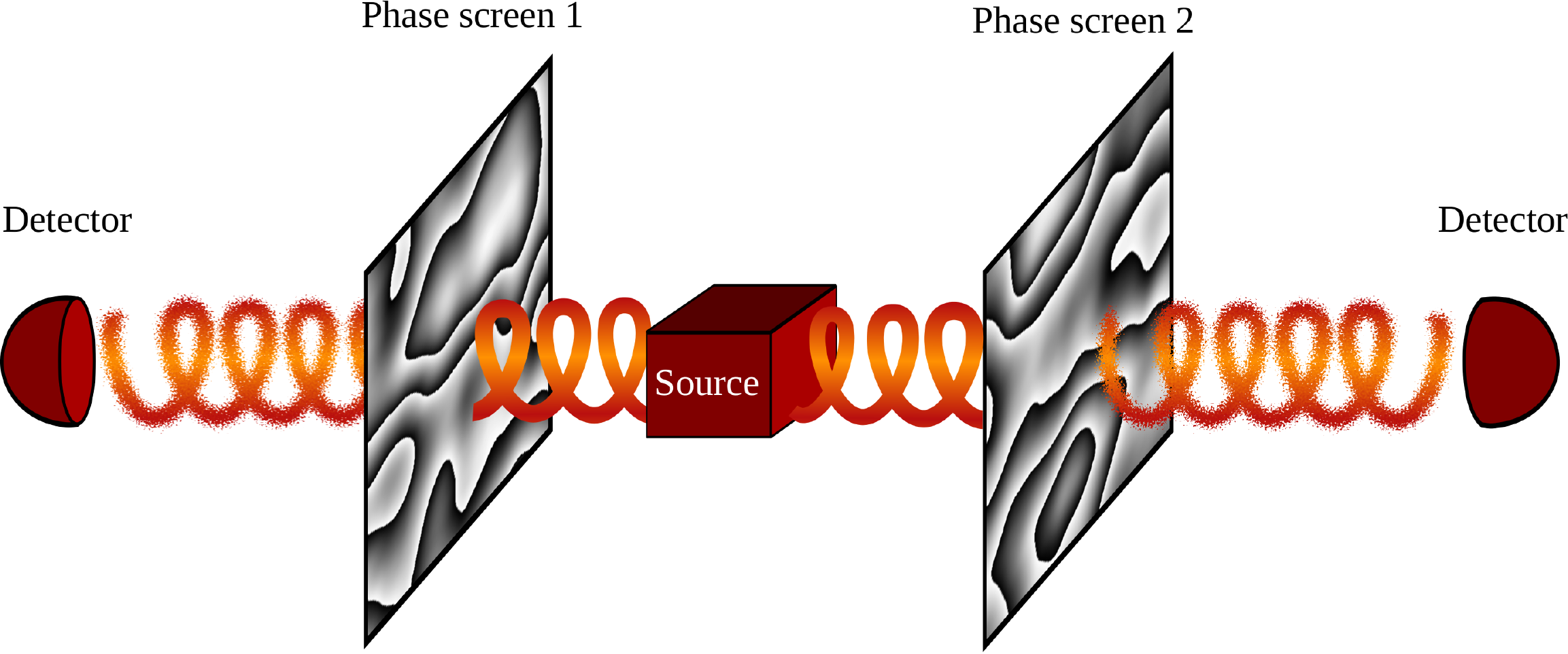}
\caption{(color online) Sketch of the setup: A source produces pairs of OAM-entangled qubits whose wave fronts get deteriorated as they propagate along the (horizontal) $z$-axis, through independent layers of a turbulent atmosphere, modeled as random phase screens. 
The characteristic scale of the screens' phase patterns is defined by the Fried parameter $r_0$, which in turn depends on the distance $L$ from source to detector (see Eq.~(\ref{Fried})).}
\label{fig:setup}
\end{figure}

We proceed with presenting our model shown in Fig.~\ref{fig:setup}. 
The source produces pairs of photons that are maximally entangled in their OAM, for example, 
by transfering the maximal entanglement of two photons in the polarization degree of freedom to the OAM degree of freedom, there encoded by LG modes with opposite azimuthal quantum number $l$ \cite{Fickler02112012}. 
We assume that the input LG modes have a waist $w_0$ (which coincides with the waist of the Gaussian TEM$_{00}$ mode \cite{andrews}), a radial quantum number $p_0=0$ and azimuthal quantum numbers $l_0$ and $-l_0$. 
The generated Bell state thus reads:
\begin{align}
|\Psi_0\ra&=\frac{1}{\sqrt{2}}\lt(|l_0,{-l_0}\ra+e^{i\gamma}|{-l_0},l_0\ra\rt),
\label{gen_Bell}
\end{align} 
where $\gamma$ is a relative phase.

The entangled OAM photons are then sent along the (horizontal) $z$-axis through independent weak turbulences modeled as phase screens \cite{paterson05}. It is convenient to assume that the screens are 
 inserted at $z=0$, in which case the phase screens act directly upon the generated state (\ref{gen_Bell}). Our fundamental quantity of interest 
will be the output density operator $\rho$ of the biphoton state upon transmission through the weakly turbulent media of thickness $|z|=L$ (for each photon). 

It is useful to recall the properties of the linear map  $\Lambda$ that represents the action of an ensemble-averaged phase screen on a single photon density operator \cite{todd_brun13}. $\Lambda$ relates the input and the output density matrices of a single photon 
in the OAM basis, $\sigma^{(0)}$ and $\sigma$, respectively, through the equation 
\be
\sigma_{pl,p^\prime l^\prime}=\sum_{p_0l_0,p_0^\prime l_0^\prime} \Lambda_{pl,p^\prime l^\prime}^{p_0l_0,p_0^\prime l_0^\prime}\sigma^{(0)}_{p_0l_0,p_0^\prime l_0^\prime},
\label{elem_sigma}
\e
and includes the propagation distance implicitly, 
through its dependence on the Fried parameter, see Eqs. (\ref{sum_p_Lambda}), (\ref{Fried}) below. The matrix elements $\Lambda_{pl,p^\prime l^\prime}^{p_0l_0,p_0^\prime l_0^\prime}$ have the following meaning: 
The ones with coinciding  indices 
$p=p_0$, $l=l_0$, $p^\prime=p^\prime_0$, $l^\prime=l^\prime_0$, describe the mapping of the initially populated OAM modes onto themselves (the ``survival amplitude''); all other matrix elements control the crosstalk to distinct modes (at least one of the indices changes its value). 

By generalizing the above description to biphotons, it is easy to show that 
the two-photon output state $\rho$ is related to the input state $\rho^{(0)}=|\Psi_0\rangle\langle \Psi_0|$ by the formula
\be
\rho=(\Lambda_1\otimes \Lambda_2) \rho^{(0)},
\label{rho_2}
\e
where $\Lambda_i$ ($i=1,2$) is a linear transformation representing the phase screens seen by either photon, respectively. In the following, we assume that the phase screens are characterized by the same statistical properties, 
which allows us to write $\Lambda_1=\Lambda_2=\Lambda$. 

We now proceed to describe the entanglement evolution under the influence of turbulence, keeping track of the azimuthal quantum number which encodes quantum information. The radial quantum number of the output state remains thereby unobserved, and is traced over. The elements of the resulting transformation, $\Lambda_{l,l^\prime}^{l_0,l_0^\prime}\equiv \sum_p \Lambda_{pl,p l^\prime}^{0l_0,0l_0^\prime}$, read:
\begin{align}
\Lambda_{l,\pm l}^{l_0,l_0^\prime}&=\frac{\delta_{l_0-l_0^\prime,l\mp l}}{2\pi}\int_0^\infty dr R_{0l_0}(r)R^*_{0l_0^\prime}(r)r\int_0^{2\pi}d\vartheta \n\\
&\times e^{-i\vartheta[l\pm l-(l_0+l_0^\prime)]}
e^{-0.5 D_\phi(2r|\sin(\vartheta/2)|)} ,
\label{sum_p_Lambda}
\end{align}
where $R_{0l_0}(r)$ is the radial part of the input LG beam at $z=0$, with radial and azimuthal quantum numbers $0$ and $l_0$, 
respectively \cite{PhysRevA.45.8185}.
$D_\phi(r)=6.88(r/r_0)^{5/3}$ is the phase structure function of the Kolmogorov model of turbulence \cite{andrews}, where
\be
r_0=(0.16 \, C_n^2k^2L)^{-3/5},
\label{Fried}
\e
is the transverse correlation length of turbulence (Fried parameter), depending on the phase structure constant $C_n^2$ \cite{andrews}, the optical wave number $k$, and the propagation distance $L$.

The linear map (\ref{sum_p_Lambda}) exhibits the inversion symmetry 
\be
\Lambda^{l_0,l_0^\prime}_{l, l'}=\Lambda^{-l_0,-l_0^\prime}_{-l,- l'},
\label{symmetry}
\e
which stems from the isotropy of turbulence along horizontal paths \cite{andrews}. 
We recall that a similar symmetry of hybrid qudits with entangled spin and OAM, reflecting their rotational invariance with respect to the propagation direction, 
was discussed in the context of alignment-free quantum communication  \cite{DAmbrosio:2012cr}. 

Due to the crosstalk, the matrix elements of the output state spread over the entire OAM basis. To deal with a finite dimensional Hilbert space, we
here consider a situation \cite{raymer06,alpha13} where the transmitted state is postselected in the basis of the injected qubit state, $\{|{-l_0},{-l_0}\ra,|{-l_0},l_0\ra,|l_0,{-l_0}\ra,|l_0,l_0\ra\}$. Since such postselection entails the decay of the output state, the decaying output biphoton state needs to be renormalized by its trace \cite{hiesmayr03} before we can quantify the entanglement evolution in turbulence by the concurrence \cite{wootters98}.

In the truncated Hilbert space that is spanned by the four two-qubit states listed above, owing to the symmetry (\ref{symmetry}), there are only two distinct non-zero matrix elements of the map $\Lambda$:
\begin{align}
a&=\Lambda_{l_0,l_0}^{l_0,l_0}=\Lambda_{{-l_0},{-l_0}}^{{-l_0},{-l_0}}=\Lambda_{{-l_0},l_0}^{{-l_0},l_0}=\Lambda_{l_0,{-l_0}}^{l_0,{-l_0}},\label{a}\\
b&=\Lambda_{l_0,l_0}^{{-l_0},{-l_0}}=\Lambda_{{-l_0},{-l_0}}^{l_0,l_0},
\label{b}
\end{align} 
which we recognize as the survival ($a$) and crosstalk ($b$) amplitudes, respectively (see our discussion following Eq. (\ref{elem_sigma})). 
By virtue of Eqs.~(\ref{rho_2})-(\ref{b}), the final state can now be analytically parametrized in terms of $a$ and $b$, and evaluated numerically, for arbitrary $l_0$. 

Then, using Wootters formula for the concurrence of a mixed bipartite qubit state \cite{wootters98}, we obtain our first result -- an analytical expression for OAM biphotons: 
\be
C(\rho)=\max\lt[0,\frac{(1-2\tilde{a})}{(1+\tilde{a})^2}\rt],
\label{conc}
\e
where $\tilde{a}\equiv b/a$. It can be seen immediately from Eq.~(\ref{conc}) that $C(\rho)=1$ for $\tilde{a}=0$, and $C(\rho)=0$ 
for $\tilde{a}\geq 1/2$. Moreover, from the definitions (\ref{a},\ref{b}) it follows that $\tilde{a}=0$ implies $b=0$, $a=1$, 
what corresponds to $r_0\to \infty$, that is, to the absence of turbulence. For finite values of the Fried parameter $r_0$, 
there emerges a non-zero crosstalk ($b>0$), which is accompanied by a decrease of the survival amplitude ($a<1$) 
and results in a monotonic growth of $\tilde{a}$ until $\tilde{a}=1/2$ (i.e., a steady decrease of $C(\rho)$ to zero).  

\begin{widetext}
\begin{center}
\begin{figure}
{\includegraphics[width=8cm]{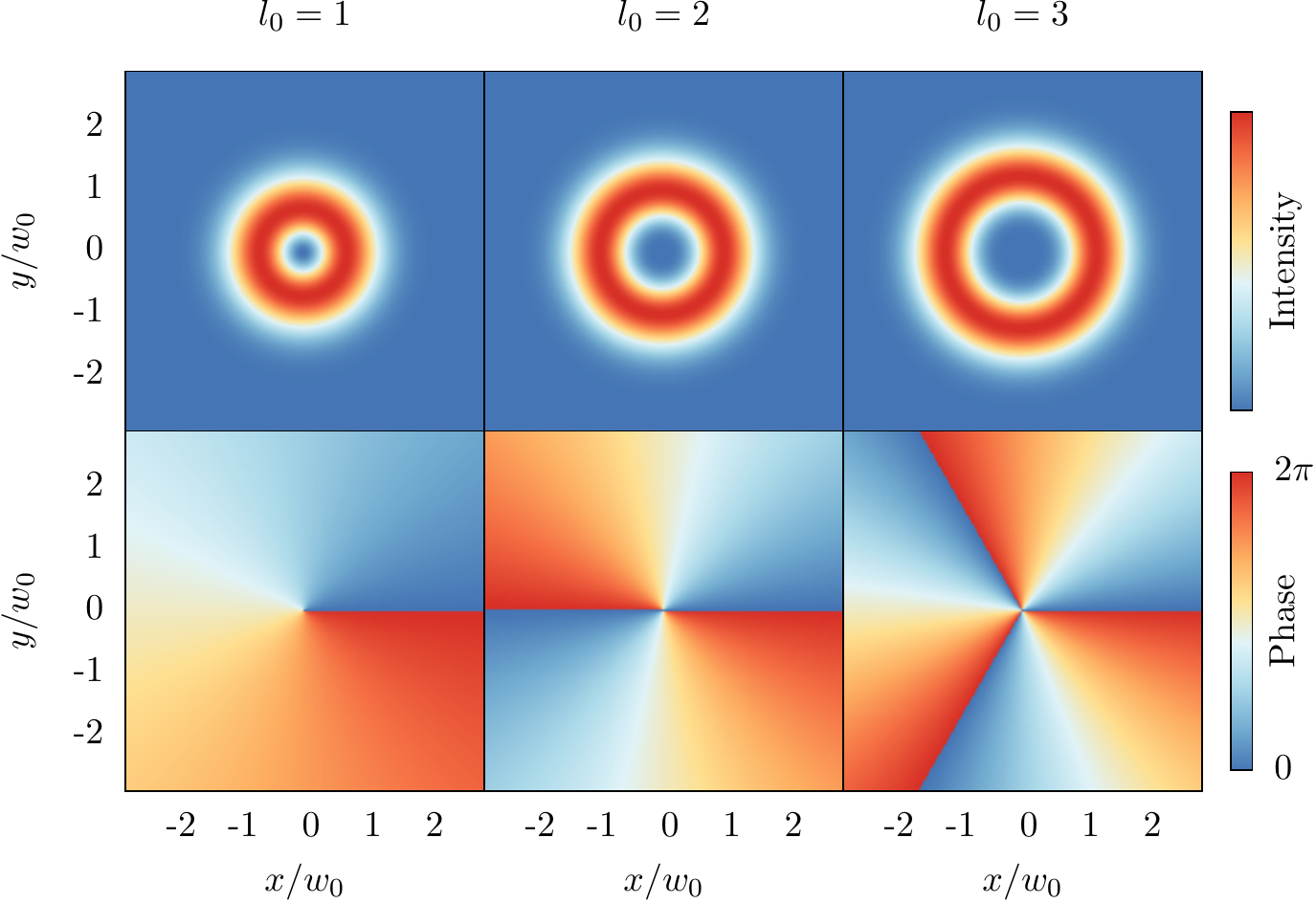}\quad\includegraphics[width=7.5cm]{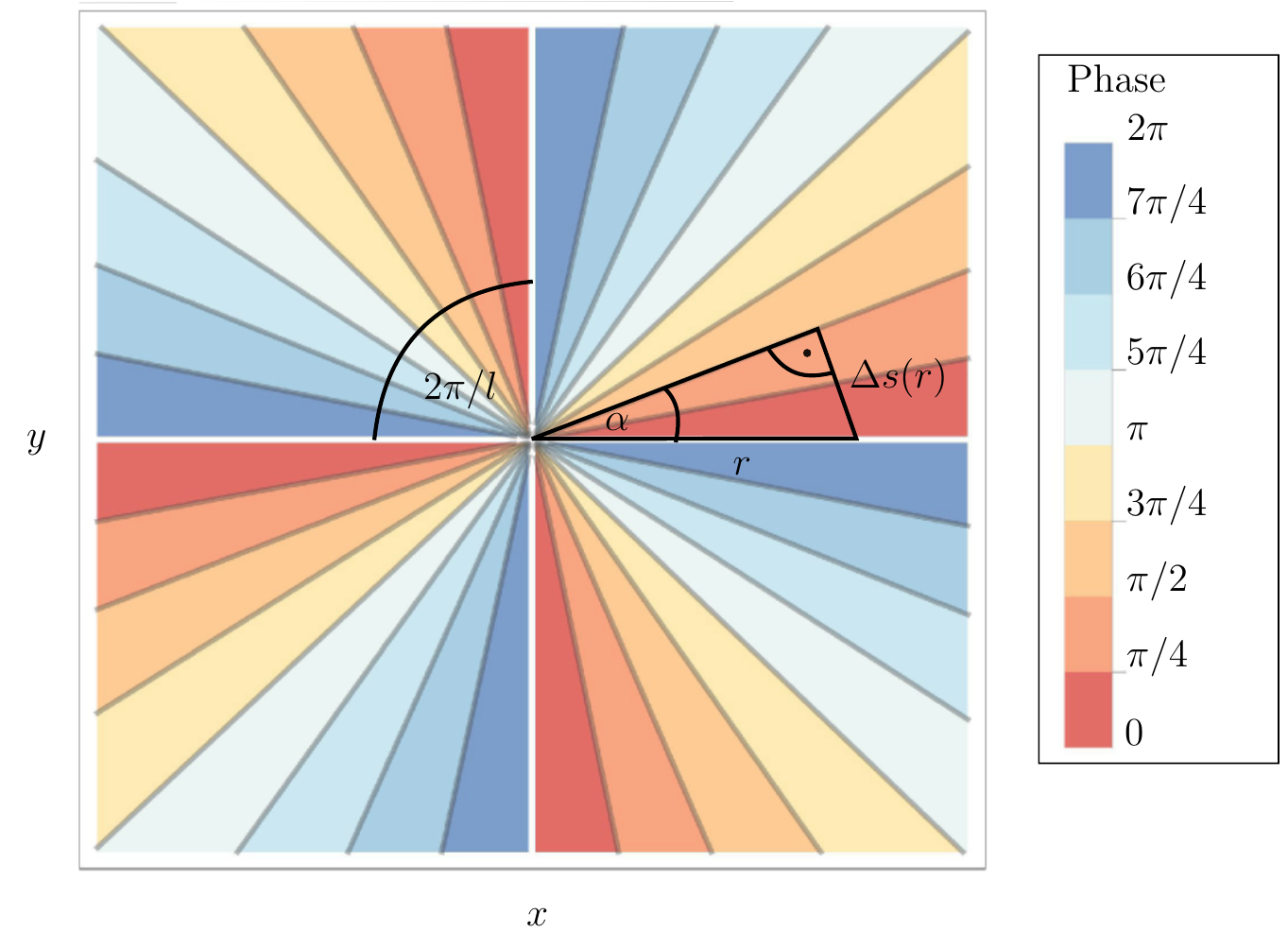}}
\caption{(color online) Illustration of the complex spatial structure of the vortex beams. Left panel (top, from left to right): Inhomogeneous intensity profile of LG beams for $l_0=1,2,3$. Note the widening of the beam with increasing $l_0$.  Left panel (bottom, from left to right): Augmented phase oscillations with increasing $l_0$, for $l_0=1,2,3$, respectively. Right panel: Sketch for the calculation of the {\it phase correlation length}. The phase variation of a LG-beam with $l_0=4$ is encoded in color with a step width of $\pi/4$. The length $\Delta s(r)$ is defined as the shortest distance from a point (here, chosen on the $x$-axis) located at a distance $r$ from the origin to the line of points with a  phase difference of $\pi/2$.}
\label{fig:structure}
\end{figure}
\end{center}
\end{widetext}
We now want to gain some insight into the physical mechanism governing the entanglement evolution of twin OAM photons in turbulence. 
Entangled qubits become more robust with increasing $l_0$ \cite{raymer06} because their spatial phase structure gets finer --
as the OAM beam widens, its phase front oscillates more rapidly with increasing $l_0$ (see Fig.~\ref{fig:structure}(bottom, left panel)). 
As a result, for a  fixed turbulence correlation length $r_0$, 
OAM-entangled qubits whose wave fronts have a shorter characteristic length than $r_0$ `see' 
turbulence as a homogeneous medium which does not affect their quantum entanglement.
When we plot $C(\rho)$ against the ratio $w_0/r_0$, this property of the wave fronts of OAM beams does implicitly come into play  as the increased longevity of the concurrence for larger $l_0$, 
since $w_0$ is $l_0$-independent \cite{raymer06,gopaul07,roux11}. However, this effect can be made strikingly obvious by a proper rescaling. 

To this end, we introduce \cite{NinaThesis} the {\it phase correlation length} $\xi(l_0)$  which we define as the average distance between 
the points in the LG beam cross-section that have a phase difference of $\pi/2$. The idea comes from the basic fact that two monochromatic waves having a phase difference $\lesssim \pi/2$  are `in phase' and interfere constructively.
From the azimuthal phase dependence of OAM beams, proportional to $e^{il_0\vartheta}$, 
it is easy to see that the angle between two such points is equal to $\alpha= \pi/2|l_0|$ [see Fig.~\ref{fig:structure}(right panel)]. 
Now, choose a point at a distance $r$ from the origin. Then the distance $\Delta s(r)$ from this point to the line along which the phase differs by $\pi/2$ from the phase of the departure point coincides with the leg of the right triangle shown in Fig.~\ref{fig:structure}(right panel), and is given by
$\Delta s(r)=r \sin\alpha=r\sin(\pi/2|l_0|)$. Also note that such a triangle cannot be constructed for $|l_0|=1$, what will manifest in a specific concurrence evolution for this case, see below.

Finally, $\xi(l_0)$ is defined as the average of $\Delta s$, or $\xi(l_0)=\langle r\rangle \sin(\pi/2|l_0|), $ 
with the intensity distribution of the initial beam as weighting function:
\be
\xi(l_0)=\int_0^\infty|R_{0l_0}(r)|^2\Delta s(r) r dr. 
\label{def_xi1}
\e
The integral in Eq.~(\ref{def_xi1}) can be evaluated exactly \cite{bronstein_book}, yielding our final expression for the phase correlation length:
\be
\xi(l_0)=\sin\lt(\frac{\pi}{2|l_0|}\rt)\frac{w_0}{\sqrt{2}}\frac{\Gamma(|l_0|+3/2)}{\Gamma(|l_0|+1)},
\label{def_xi}
\e
where $\Gamma(x)$ is the Gamma function. For $|l_0|>2$, the function $\xi(l_0)$ is monotonically decreasing with $|l_0|$, what is consistent with the faster phase oscillations of LG beams for larger $|l_0|$. 

\begin{figure}
\includegraphics[width=6.5cm]{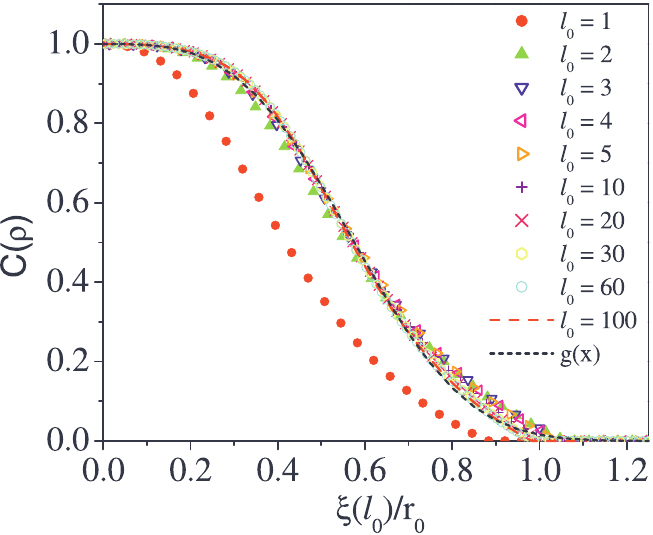}
\caption{(color online) Output state concurrence (\ref{rho_2}) as a function of the ratio $\xi(l_0)/r_0$, for different $l_0$. Concurrences for $l_0>1$ essentially collapse onto a universal, exponential decay law which is best fitted by $g(x)= \exp(-4.16 x^{3.24})$.}
\label{uni_conc}
\end{figure}
In Fig.~\ref{uni_conc} we plot our numerical results for the concurrence $C(\rho)$ (Eq.~(\ref{conc})) as a function of the ratio 
$x\equiv \xi(l_0)/r_0$, and for different values of $l_0$: Apart from a finite size effect for $|l_0|=1$, the output state entanglement for all variable-$l_0$ initial states collapses onto one universal curve, $C(\rho)\approx \exp(-4.16 x^{3.24})$, where the exponential fit is obtained from the $x$-dependence of $C(\rho)$ for $l_0\gtrsim 50$, when all curves become indistinguishable for different $l_0$-values. The thus demonstrated universality of OAM entanglement decay in turbulence is the key result of our present contribution. It shows that the entanglement evolution of OAM qubit states in turbulence is governed by the sole parameter $\xi(l_0)/r_0$. It should be mentioned that a different rescaling was done in \cite{paterson05}, where only the broadening, but not the phase oscillations, of the LG beam with increasing $l_0$ was taken into account. Therefore, such rescaling cannot unveil the here uncovered universality of the concurrence decay. 
Using Fig.~\ref{uni_conc}, definition (\ref{def_xi}), and recalling the properties of the Gamma function \cite{handbook}, we obtain an asymptotic ($l_0\to\infty$) scaling law, $L\sim |l_0|^{5/6}$, for the dependence of the propagation distance $L$ (over which the concurrence remains finite) on $l_0$. We thus analytically derive an earlier, phenomenological result of \cite{alpha13}.

To conclude, we have introduced the phase correlation length $\xi(l)$ of OAM beams and shown that it fully determines the entanglement evolution of OAM qubit states in weak turbulence. Since $\xi(l)$ reflects the complex spatial structure of OAM beams and is independent of the turbulence model, it will be interesting to apply this quantity to study entanglement evolution in strong turbulence. Another direction of future work will be to see whether a generalization of the phase correlation length to high dimensions -- as a weighted sum of partial phase correlation lengths of individual OAM components -- can be useful for an improved understanding of the entanglement evolution of OAM qudit states in turbulence and/or for the identification of high dimensional and {\it robust} entangled OAM states.

\bibliographystyle{h-physrev} 
\bibliography{OAM_31}

\begin{thebibliography}{10}

\bibitem{leach02}
J.~Leach, M.~J. Padgett, S.~M. Barnett, S.~Franke-Arnold, and J.~Courtial,
\newblock Phys. Rev. Lett. {\bf 88}, 257901 (2002).

\bibitem{vaziri03}
A.~Vaziri, J.-W. Pan, T.~Jennewein, G.~Weihs, and A.~Zeilinger,
\newblock Phys. Rev. Lett. {\bf 91}, 227902 (2003).

\bibitem{zeilinger06}
S.~Gr\"oblacher, T.~Jennewein, A.~Vaziri, G.~Weihs, and A.~Zeilinger,
\newblock New J. Phys. {\bf 8}, 75 (2006).

\bibitem{Molina-Terriza:2007zr}
G.~Molina-Terriza, J.~P. Torres, and L.~Torner,
\newblock Nat. Phys. {\bf 3}, 305 (2007).

\bibitem{Nagali:2009qf}
E.~Nagali {\em et~al.},
\newblock Nat. Photon. {\bf 3}, 720 (2009).

\bibitem{pors11}
B.-J. Pors, F.~Miatto, G.~W. 't~Hooft, E.~R. Eliel, and J.~P. Woerdman,
\newblock J. Opt. {\bf 13}, 064008 (2011).

\bibitem{leach12}
J.~Leach, E.~Bolduc, D.~J. Gauthier, and R.~W. Boyd,
\newblock Phys. Rev. A {\bf 85}, 060304 (2012).

\bibitem{Fickler02112012}
R.~Fickler {\em et~al.},
\newblock Science {\bf 338}, 640 (2012),
  http://www.sciencemag.org/content/338/6107/640.full.pdf.

\bibitem{Wang:2012uq}
J.~Wang {\em et~al.},
\newblock Nat. Photon. {\bf 6}, 488 (2012).

\bibitem{Torres:2012zr}
J.~P. Torres,
\newblock Nat. Photon. {\bf 6}, 420 (2012).

\bibitem{paterson05}
C.~Paterson,
\newblock Phys. Rev. Lett. {\bf 94}, 153901 (2005).

\bibitem{Pors:11}
B.-J. Pors, C.~H. Monken, E.~R. Eliel, and J.~P. Woerdman,
\newblock Opt. Express {\bf 19}, 6671 (2011).

\bibitem{Malik:12}
M.~Malik {\em et~al.},
\newblock Opt. Express {\bf 20}, 13195 (2012).

\bibitem{alpha13}
A.~Hamadou~Ibrahim, F.~S. Roux, M.~McLaren, T.~Konrad, and A.~Forbes,
\newblock Phys. Rev. A {\bf 88}, 012312 (2013).

\bibitem{PhysRevA.88.053836}
M.~V. da~Cunha~Pereira, L.~A.~P. Filpi, and C.~H. Monken,
\newblock Phys. Rev. A {\bf 88}, 053836 (2013).

\bibitem{rodenburg14}
B.~Rodenburg {\em et~al.},
\newblock New J. Phys. {\bf 16}, 033020 (2014).

\bibitem{raymer06}
B.~J. Smith and M.~G. Raymer,
\newblock Phys. Rev. A {\bf 74}, 062104 (2006).

\bibitem{gopaul07}
C.~Gopaul and R.~Andrews,
\newblock New J. Phys. {\bf 9}, 94 (2007).

\bibitem{roux11}
F.~S. Roux,
\newblock Phys. Rev. A {\bf 83}, 053822 (2011).

\bibitem{Sheng:12}
X.~Sheng, Y.~Zhang, F.~Zhao, L.~Zhang, and Y.~Zhu,
\newblock Opt. Lett. {\bf 37}, 2607 (2012).

\bibitem{todd_brun13}
J.~R. Gonzalez~Alonso and T.~A. Brun,
\newblock Phys. Rev. A {\bf 88}, 022326 (2013).

\bibitem{Mintert05b}
F.~Mintert, A.~R.~R. Carvalho, M.~Ku\'{s}, and A.~Buchleitner,
\newblock Phys. Rep. {\bf 415}, 207 (2005).

\bibitem{tiersch08}
M.~Tiersch, F.~de~Melo, and A.~Buchleitner,
\newblock Phys. Rev. Lett. {\bf 101}, 170502 (2008).

\bibitem{hiesmayr03}
R.~A. Bertlmann, K.~Durstberger, and B.~C. Hiesmayr,
\newblock Phys. Rev. A {\bf 68}, 012111 (2003).

\bibitem{andrews}
L.~C. Andrews and R.~L. Phillips,
\newblock {\em Laser Beam Propagation through Random Media}, {second} ed. (SPIE
  Press, Bellingham, 2005).

\bibitem{PhysRevA.45.8185}
L.~Allen, M.~W. Beijersbergen, R.~J.~C. Spreeuw, and J.~P. Woerdman,
\newblock Phys. Rev. A {\bf 45}, 8185 (1992).

\bibitem{DAmbrosio:2012cr}
V.~D'Ambrosio {\em et~al.},
\newblock Nat. Comm. {\bf 3}, 961 (2012).

\bibitem{wootters98}
W.~K. Wootters,
\newblock Phys. Rev. Lett. {\bf 80}, 2245 (1998).

\bibitem{NinaThesis}
N.~Leonhard,
\newblock Propagation of orbital angular momentum photons through atmospheric
  turbulence,
\newblock Diploma thesis (Albert-Ludwigs University of Freiburg), 2014.

\bibitem{bronstein_book}
I.~N. Bronstein and K.~A. Semendyayew,
\newblock {\em Handbook of Mathematics} (Harri Deutsch, Frankfurt am Main,
  1985).

\bibitem{handbook}
M.~Abramowitz and I.~Stegan, editors,
\newblock {\em Handbook of Mathematical Functions with Formulas} (National
  Bureau of Standards, Washington, D.C., 1964).

\end{thebibliography}

\end{document}